\documentclass[preprint,showpacs]{revtex4}

\usepackage{graphicx}

\begin{document}

\title{Classical Inversion of the CHSH Inequality}

\author{Frederick H.~Willeboordse}
\homepage{http://chaos.nus.edu.sg/}
\email{willeboordse@yahoo.com}
\affiliation{University Scholars Programme, The National University of Singapore, Singapore 117542}

\date{\today}

\begin{abstract}
In general, the CHSH inequality $S \le 2$ is considered to provide an \textit{upper} bound for classical correlations. In this note it is shown that if incoming particles are allowed to be delayed ever so briefly, the inequality can be inverted to $S \ge 2$ thus becoming a \textit{lower} bound for classical correlations (maximum $S=3$). All interaction is strictly local and events are neither dropped nor re-sequenced thus strictly conforming to the standard CHSH setup. The key notion is that a (brief) delay in front of  a detector is not a causative intervention.
\end{abstract}

\pacs{03.65.Ud}
\keywords{Bell inequalities, local realism}

\maketitle

\begin{figure}[htb]
   \includegraphics*[width=8cm]{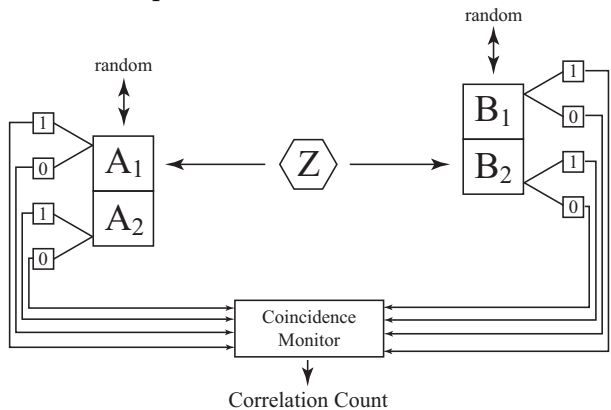}
   \caption
   {
     Typical setup in a CHSH style Bell correlation experiment. The source Z sends out a pair of entangled particles toward detectors A and B that can randomly be in 
     position 1 or 2. The binary output of the detectors is collected by a coincidence monitor. 
   }
   \label{figure1}
\end{figure}

The Clauser, Horne, Shimony and Holt (CHSH)\cite{CHSH} inequality is used commonly in discussions concerning the foundations of quantum mechanics, secure communication and quantum information theory. A typical experimental setup is depicted in Figure \ref{figure1}. In this setup, a source creates an entangled pair of particles (usually photons) that interact with a detector that can be in one of two possible settings (e.g.~the direction with respect to which polarization is measured). The detector setting must be completely random, unknown to the source at the time a pair of particles is created, and independent of the other detector's setting. Consequently, communication between the detectors, the source and the detectors etc.~is not allowed thus ensuring that the physical processes occurring when a particle hits a detector are local. When a particle interacts with a detector, a binary outcome is generated which together with the detector setting is either forwarded to a coincidence monitor or recorded with accurate clocks for later comparison.

\vspace{-3mm}

One can then determine the various (anti-)correlations between the right and left hand detector settings and compare the experimental outcomes with quantum mechanical and classical calculations. Denoting the expectation values for the (anti-)correlations as $E(A_i,B_j)$ where the detectors $A$ and $B$ have settings $i,j = 1,2$ respectively, CHSH consider the following sum of (anti-)correlations \cite{CHSH}
\begin{equation}
  S = E(A_1,B_1) + E(A_1,B_2) + E(A_2,B_1) - E(A_2,B_2).
	\label{CHSH-Ineq}
\end{equation}
Quantum mechanically, the maximum is $S = 2\sqrt{2}$ while in a realistic model, the maximum is generally argued to be $S = 2$ \cite{Shimony_Stanford}. The standard realistic result is easily illustrated: assume that the photon carries the binary information for each possible detector setting. For example, if we take $A_1=1$, $A_2=0$, $B_1=0$ and $B_2=0$ (in total there are 16 possibilities), then $S = 2$; changing a 0 to a 1 may at best improve the (anti-)correlation with one detector but this will be at the expense of the (anti-)correlation with another detector so that the maximum remains $S=2$.

However, what has not adequately been considered is the possibility for the incoming particle to briefly be delayed in front of the detector based on the detector setting. Such a delay and observation does not contravene any of the standard requirements of a CHSH experiment. Notably, it does not imply non-local communication, the dropping of particles or events or non-perfect randomness of the detector settings. A brief delay depending on the encountered detector setting is a perhaps unlikely but nevertheless physically conceivable. For example, if the detector setting were the orientation of a rectangular slit inside some box with cleverly arranged tracks, then an incoming cuboid particle might go right through if it is aligned with the slit but move around on the tracks for some time if it is not aligned with the slit.

\begin{figure}[htb]
   \includegraphics*[width=8cm]{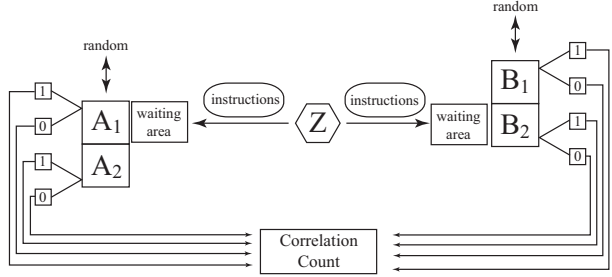}
   \caption
   {
     A classical CHSH apparatus for which $S \ge 2$. There is no coincidence monitor since each particle leads to exactly one output, and the outputs are processed in sequence. I.e.~one particle pair leads to an output from each detector and the (anti-)correlation of the outputs is immediately recorded (all events are recorded and nothing is dropped or added).
   }
   \label{figure2}
\end{figure}

\begin{figure}[htb]
   \includegraphics*[width=6cm]{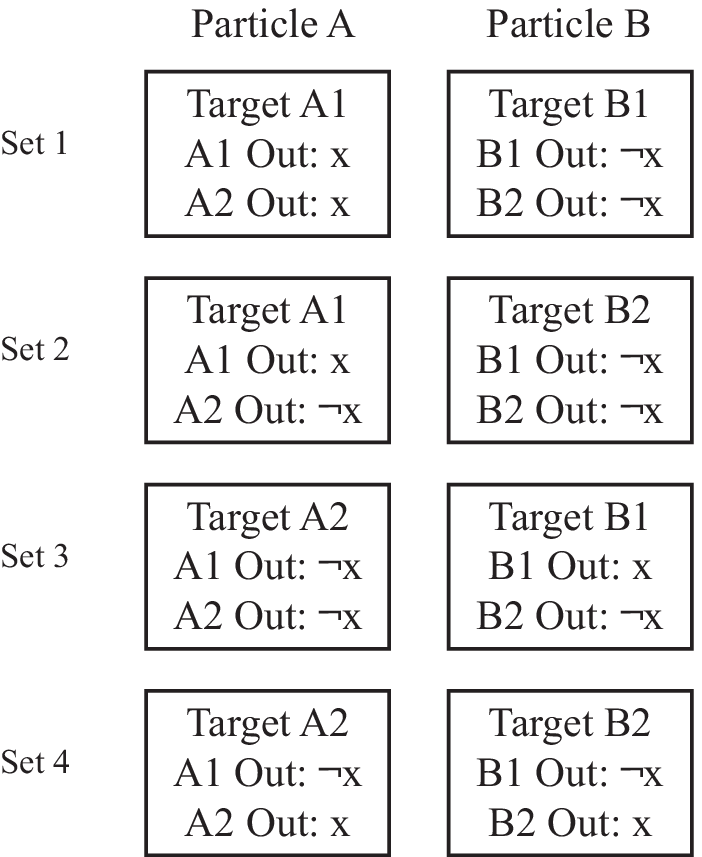}
   \caption
   {
     Instructions set for the particles. $x$ is a uniformly randomly chosen boolean value so that for each detector setting the probability to have 0 or 1 as the output is 50\%.
   }
   \label{figure3}
\end{figure}

It will now be shown that by allowing a delay and local observation, given a correctly designed instruction table, the CHSH inequality can be reversed. To do so, the classical apparatus shown in Figure \ref{figure2} is considered which is very similar in setup to Figure \ref{figure1}. The apparatus functions as follows:
\begin{itemize}
  \item A source creates two particles A and B, and uniformly randomly assigns each particle a target detector setting $A_{\tau}$, $\tau = 1,2$. At the source, the particles receive a set of instructions that determine the output generated when they reach the detector. There are four different instruction sets corresponding to the randomly selected target settings, see Figure \ref{figure3}. The particles do not need to know each others instructions sets. There is however, a dependence on the other particle's target. It must be stressed that this does not contravene locality since the instruction sets are given to the particles at the source immediately after the target settings are assigned.
  \item The particles fly toward the detectors. The particles are not allowed to communicate with each other or the source (or anything else for that matter).
  \item The two detectors are set by a random process, independent of each other and without any form of communication.
  \item When a particle arrives at a detector, it inspects the detector's setting (this is allowed since it is a local interaction). If the setting is the target setting $A_{\tau}$, the particle generates an output according to the instructions it has carried along. Otherwise it is delayed.
  \item If the particle is delayed in front of the detector, it remains there for one (or possibly more) time step(s) and then generates an output according to instructions it has carried along.
\end{itemize}

\vspace{-8mm}

It is straightforward for the apparatus to yield $S=2$ by giving the particles the right instructions for the output such as for example $A_1=0$, $A_2=1$, $B_1=0$ and $B_2=1$. Indeed, one can construct instructions such as those in Figure \ref{figure3} so that if either both particles or one of the particles hits a target detector setting, the output will be the (anti-)correlation as desired. With an instruction set such as in Figure \ref{figure3}, the only configuration that yields the 'wrong' (anti-)correlations is when both particles do \textit{not} hit the target setting (here, 'wrong' (anti-)correlation is used for an (anti-)correlation that decreases $S$ from the maximum value $S=3$). When there is no delay, the probability that both particles do not hit their target setting is 25\% so that $S=0.75+0.75+0.75-0.25=2$, in accordance with the regular CHSH inequality.

Now if there is a delay, then there is a finite probability that either one or both the detectors switch during the delay. Consequently, the wrong (anti-)correlation only occurs when neither detector switches. The probability that this happens is given by $A = \frac{1}{4} (1-p)^2 $, where $p$ the (local) individual probability that a detector switches. In the arrangement used in Eq.~\ref{CHSH-Ineq}, for detector settings A1-B1, A1-B2, A2-B1, the fraction of anti-correlations then becomes $1-A$, while for A2-B2 the correlations equal $A$. Consequently, given the instruction table in Figure \ref{figure3}, when delaying the output for one time step the CHSH inequality becomes
\begin{equation}
   S = 3 - (1 - p)^2 \ge 2
\end{equation}
which is the inverse of the usual case.

\vspace{-4mm}

While the described method clearly fulfills all the requirements for a Bell setup, an area that may be of contention is the legitimacy for an arriving particle to be delayed if it doesn't encounter the setting it is targeted at. Wouldn't that be the same as simply asking the particle to actively set the detector? (It is trivial that if the particle is allowed to set the detector in a certain position that $S=3$). This is not the case, delaying an action while waiting for an event is not the same as causing that event. For example, if I delay my wake-up time until the sun rises, it doesn't cause the sun to rise. Clearly, waiting is a legitimate local action. The inverted inequality therefore holds. It's dependency on the switching frequency is physically unsatisfactory and unsupported by experiment. However, the result does show that violation of the usual inequality is insufficient to establish an exclusive quantum domain.

\bibliography{bsng}

\end{document}